# General Relativity Constraints the Proper Times and Predicts the Frozen Stars Instead of the Black Holes

Zahid Zakir

*Centre for Theoretical Physics and Astrophysics,*
*11a, Sayram, 100170 Tashkent, Uzbekistan*
*ctpa@theorphys.org*

**Abstract**

In a static gravitational field an intersection of a worldline by a global hypersurface of simultaneity t=const gives an invariant constraint relating the proper time of this event by t. Since at any finite t the such constrained proper time intervals are less than reqired for crossing a horizon, general relativity predicts the gravitational freezing of proper times in stars with time-like or null geodesics everywhere. The time dilation stabilizes contracting massive stars by freezing, which is maximal but finite at the centre, and the surface is frozen near the gravitational radius. The frozen stars (frozars) slowly defrost due to emissions and external interactions, the internal phase transitions can initiate refreezing with bursts and explosions.

1. **Introduction**

The massive cold stars cannot support themselves against the gravitational contraction. In the Newtonian theory this leads to the quick collapse into the region $r_g = 2GM$. In general relativity (GR) the surface of the star becomes frozen near the gravitational radius $r_g$ for the distant observers.

But when the surface contracts, the proper time $\tau_g$ needed to cross $r_g$ remains finite and the theory must to interpret this fact. On the basis of this fact the *black-hole* approach predicts the real *black holes* (BHs) [1-2]. But one of the basic implicit assumptions in this approach is that the evolution in terms of the proper times can be considered independently from the coordinate time's moments.

Recently in the paper [3] it has been proposed a new treatment of the massive relativistic stars using the fact that in the static fields GR allows one to define a global simultaneity of events in terms of the coordinate time *t*. It has been shown that GR allows one to construct on the hypersurfaces of simultaneity *t=const* a new theory of compact massive stars as the gravitationally frozen objects (frozars) having some internal structure at any fixed moment *t*.

In the present paper it will be shown that in GR the *identity* for the radial coordinate of an event $r(t) = r(\tau)$ leads to an *invariant constraint* for the proper times of falling particles $\tau = \tau(t)$ restricting them by relating with *t*. Some new observable properties of the such frozen stars (frozars) predicting by GR will be discussed briefly.

**1   The physical coordinates and a global simultaneity in a static field**

In a rest frame of a centrally-symmetric source the spherical symmetry leads to three key physical properties of the gravitational field *around* that source: (1) *it is static;* (2) *it does not change the length of a standard rod oriented perpendicularly to radius,* and (3) *there is a global simultaneity of events on the hypersurfaces* $t = const$ *defined via a set of coordinate clocks synchronized by distant observer's clocks*.

In most of coordinate systems of various frames of reference these physical properties become *hidden* due to insufficient geometrical or kinematical complications. The curvature coordinates $(t, r, \theta, \varphi)$ of a static frame, defined on the hypersurfaces of simultaneity $t = const$, in a simplest and physically convenient form express these physical properties and the symmetries of the *static* field. In terms of the curvature coordinates the metric outside a spherical body $r > R$ (here $R > r_g$) is the Schwarzschild vacuum solution, the metric inside the source ($r < R$), as the matter solution, is regular at any point of the body.

The gravitational time dilation defining by the component of the metric $g_{00}(r) = 1 - r_g/r$ at *first* leads to the well known gravitational redshift (blueshift) of outfalling (falling) photons which in the lowest order is the same as in the Newtonian theory.

But *the second* observable effect *directly* showing the time dilation of the standard clocks at comparison with the distant observer's clocks after long term duration in the field, is not so known, but is more important since *it is absent in the Newtonian theory*. Really, in the Newtonian gravity the rate of a standard clock does not depend on the duration time $T$ in the field. In GR the dilation $\delta\tau$ of a standard clock in the gravitational field with respect to the same clock at the height $h$ is proportional to the duration $T$ and is given by the formula: $\delta\tau = g\,hT$ (here $g$ is the gravitational acceleration). This integral effect is confirmed in experiments by the clocks long-term traveled in airplanes and satellites [4-6]. Thus we must strongly follow to Einstein's treatment of the proper times as really dilated in the gravitational fields (see [6]).

In GR, due to the different rate of the standard clocks at various points, one can introduce an additional set of the *coordinate clocks* which regulated to go promptly than the standard clocks so that always *be synchronous by distant observer's clocks* (see [2]) and at any point show the coordinate or world time $t$. Therefore, any event at $r$ is labeled not only by a proper time $\tau(r)$, but by a globally defined time $t$ of a nearby coordinate clock. The such defined global hypersurface $t = const$ intersects any worldline in the field to three parts containing the events *before*, *simultaneous* and *after* the distant observer's proper time moment *t*. Although at the coordinate transformations the hypersurface can be deformed, nevertheless, the separation of events to these three groups remains unchanged due to the invariant character of the worldlines.

Thus, in GR the relative rates of the standard clocks depend on their location with the absolute dilation of the proper time of a clock in the stronger field, and this, experimentally checked [4-6] dependence is invariant fact with which agree all observers. Notice, that the existence in GR of the globally synchronized set of coordinate clocks rejects the arguments about difficulties by the informing of the distant observers since a world time moment *t* of any event can be determined instantaneously via the nearby coordinate clock. Therefore, if the equations of motion of GR require $t_g \to \infty$ for the crossing $r_g$, then this event really occurs when the *local* coordinate clocks near the source show $t_g \to \infty$.

Notice also that the observers near the massive star detect the violet-shift of photons emitted by the distant observer and they are agree that distant observer's clocks are going *faster* than the nearby standard clocks. This means that the time dilation in the gravitational field, as in the case of the twin effect in the accelerated frames, is *irreversible* and *asymmetric* under the observers. This kind of time dilation is in principle different than the *symmetric* under the observers and *relative* time dilation in the case of inertial frames on flat space-time. Thus, the approaches treating the gravitational time dilation as a relative effect are based on a wrong analogy with the time dilation on the inertial frames.

It is widely accepted also an *assumption* that the curvature coordinates in the static field have not direct physical meaning and can be replaced by more convenient ones. Really, most of arbitrary coordinates contain the "coordinate effects" not describing the properties of the underlying manifold. However, if a coordinate system is constructed as a physical one, i.e. if the coordinates mean the numbers of *resting* with respect to each other standard rods and the indications of the *globally synchronized* clocks, then the corresponding metrics describe the physical properties of the manifold. The curvature coordinates are the such physical coordinates on the spheres around the source. The corresponding Schwarzschild's metric, therefore, describes the true *physical properties* of the static field. Really, at $r \to r_g$ the proper times really dilate, the radial rods really contract with respect to the flat case, and these effects are experimentally checked. Therefore, the ignoring the such physical character of the curvature coordinates is in fact incompatible with GR and experiments.

## 2  A constraint for the proper times of particles and a falling thin dust shell

Let us consider a thin dust shell radially falling in own gravitational field. The metric outside and on the sphere $r \geq R$ is the Schwarzschild metric, the interior metric components are constant. For the particles of the shell freely falling from the rest at the spatial infinity, at a finite $r$ their coordinate velocity is given by $v(r) = dr/dt$. If at any large but finite radial coordinate $r_1 \gg r_g$ one take zero initial times: $\tau_1 = t_1 = 0$, then at $r < r_1$ one has the standard GR equation of motion:

$$t(r) = C(r_1) - 2(rr_g)^{1/2} - \frac{2}{3}\frac{r^{3/2}}{r_g^{1/2}} + r_g \ln\left|\frac{r^{1/2} + r_g^{1/2}}{r^{1/2} - r_g^{1/2}}\right|, \qquad (1)$$

where $C(r_1) = const$ is chosen so that $t(r_1) = 0$. The same world-line, parametrized across the proper time, can be described also by the equation of motion:

$$\tau(r) = \frac{2}{3r_g^{1/2}}\left(r_1^{3/2} - r^{3/2}\right), \qquad r(\tau) = \left(r_1^{3/2} - a\tau\right)^{2/3}, \qquad (2)$$

where $a = 3r_g^{1/2}/2$.

Since two forms of the equations of motion (1) and (2) describe the same particle at the same point $r$ in terms of two kind of times, they relate any moment $\tau(r)$ with corresponding global time moment $t(r)$ through the value $r$ as a parameter. Therefore, the *identity*:

$$r(t) = r(\tau) \qquad (3)$$

is in fact a *constraint*, strongly relating the proper times of the events on the worldline by the coordinate time moments $\tau = \tau(t)$. By inserting the value $r$ from (2) into (1) one obtains a direct dependence between two times $t(\tau)$ (Fig. 1):

$$t(\tau) = C(r_1) - 2r_g^{1/2}\left(r_1^{3/2} - a\tau\right)^{1/3} - \frac{2}{3r_g^{1/2}}\left(r_1^{3/2} - a\tau\right) + r_g \ln\left|\frac{\left(r_1^{3/2} - a\tau\right)^{1/3} + r_g^{1/2}}{\left(r_1^{3/2} - a\tau\right)^{1/3} - r_g^{1/2}}\right|. \qquad (4)$$

The particles approach the gravitational radius $r \to r_g$ only at $t \to \infty$ while the corresponding proper time moment $\tau_g$ remains finite. But at any finite moment of world time $t < \infty$ one has $r(t) > r_g$ and the proper time $\tau(r)$ of the particles will be less than time $\tau_g$ required for the reaching the horizon:

$$\tau_g - \tau(t) = \frac{2}{3r_g^{1/2}}\left(r^{3/2}(t) - r_g^{3/2}\right) > 0, \qquad \tau_g = \frac{2}{3r_g^{1/2}}\left(r_1^{3/2} - r_g^{3/2}\right) \qquad (5)$$

At $r \to r_g$ one has approximately $r = r_g + h$, where $h \ll r_g$, and then the equations for the trajectory and the relations between two times become very simple:

$$\tau \simeq \tau_g - h, \qquad t \simeq C' + r_g \ln\left|\frac{2r_g}{h}\right| = C' + r_g \ln\left|\frac{2r_g}{\tau_g - \tau}\right|, \qquad (6)$$

where $C' = C(r_1) - 8r_g/3$. By expressing $h$ as: $h \simeq \tau_g - \tau = b\exp(-t/r_g)$, where $b = 2r_g \exp(C'/r_g)$, one obtains an *invariant* form of the constraint for $\tau(t)$ near $\tau_g$:

$$\tau = \tau_g - be^{-t/r_g} < \tau_g, \qquad (7)$$

Thus, at any $t < \infty$, one has $\tau(t) < \tau_g$, i.e. the constraint (7) restricts the proper time's rate by the world time rate on the globally defined hypersurfaces of simultaneity (Fig.1.). Since the latter (with insufficient corrections due to the local velocities) is nothing but as the mean evolution time rate of the Universe (in a rest frame of CMB), and the intervals of $t$ are equidistant for most of the matter of the Universe, the proper times near the sources in fact are absolutely slowered with respect to this global time.

Thus, for any observer on the global hypersurface $t = const$ the dust shell does not cross own gravitational radius ($r(t) > r_g$) and the transition to the proper times or any other parametrization does not change this fact.

Since the surface of a sphere does not reach $r_g$ at any finite $t$, *inside* the shell, where the metric is homogeneous and the time dilation coefficient is the same as on the surface, the event horizon also does not arise.

Therefore, for the particles of non-zero mass the spacetime interval is timelike $ds^2 > 0$ everywhere $r \geq 0$, and the temporal and radial components of the metric remain regular both outside and inside the shell.

The same properties will have a system of enclosed spherical dust shells, and their free falling will not lead to the formation of the horizon neither outside, nor inside the such system of shells. Really, any shell changes the time dilation coefficients $g_{00}^{-1}(R_n)$ for all others to a finite value and at a finite mass it remains finite at any point: $1 \leq g_{00}^{-1}(R_n) < \infty$.

In the case of charged and rotating sources of the stationary gravitational field there are also two forms of equations of motion in terms of the world time or the proper time of falling test particles. These equations again give one a constraint for the proper times $\tau(t)$ so that at $t < \infty$ the surface will be outside the region where the time dilation coefficient tends to infinity.

### 3  General properties of the gravitationally-frozen stars

Contracting massive stars with surface at $R > r_g$ may be described in the first approximation as a set of enclosed dust shells of radius $r \leq R$. Let us consider general properties of the such stars becoming gravitationally-frozen at $R = r_g + h$, $h \ll r_g$ due to the large time dilation effects in entire volume, maximal at the centre and minimal on the surface. Their characteristic properties can be joined to the following groups [3]:

1. The large time dilation on the surface leads to the large redshift of the emitted radiation. The radiated energy is concentrated on the long-wave region of the spectrum, and the such stars should be powerful radio sources. The redshift of lines of massive stars, nuclei of galaxies and quasars, in addition to the cosmological redshift, contain also the contribution of the gravitational redshift, and new methods for the separation of the contribution each of these mechanisms should be found. It is also important that some of objects catalogued as quasars may by the Milky Way stars the redshift of which has a pure gravitational origin. Since for the extragalactic compact sources only a part of their redshift will have a gravitational origin, after the subtracting of that contribution, a pure cosmological redshift becomes less than the observing one. The such reducing of redshifts of high redshift quasars leads also to the reducing of their brightnesses, distances and masses.

2. The large time dilation near the surface also leads to the lack of sharp changes of parameters and very smooth, lengthened character of any peaks of intensity of the radiation. Particularly, this leads to a special character of the surface collisions of matter due to the time dilation in the form of a soft braking on the surface.

3. High speed rotation due to the compactness leads, particularly, to the Doppler widening of the spectral lines of radiation from the atmosphere, and also to the non-uniform distribution of charges on the equator and poles which can generate powerful magnetic fields.

4. The disks, beams and other processes around the such stars can be directly observed.

5. The quantum processes on these strong fields lead to the creation of particles. At the vacuum fluctuations of quantum fields one of particles of a created pair can fall and will be frozen on the star, while other can leave the star. In terms of $t$ the leaving particles should be described via the density matrix containing mixed states of wave functions at different moments of proper times of quanta. The such partial loss of information because of the freezing a part of quantum fluctuations near the surface leads to the occurrence of an entropy and a temperature.

6. The frozen stars have higher local temperatures than the neutron stars, and due to the interactions of practically bordering hadrons they become powerful sources of photons and neutrino. The energy loss due to various such effects, the bursts and explosions at internal phase transitions (also dilated in time) lead to slowly or rapidly "defrosting" of the frozen star.

7. The theory of quasars and galaxy nuclei, allowing one to understand their origin, now can be developed on a new fundamental basis. The gravitationally-frozen matter will be superdense mainly in compact central parts of quasars and AGNs. In the closer to a surface layers the matter density may be ordinary, although may be highly frozen in terms of $t$. Therefore, the such objects have a complicated structure with matter in various phase states.

8. Some of "candidates for BHs" are more silent than the neutron stars and the thermonuclear explosions at falling of matter to the surface have not been observed. Some compact supermassive objects at the nucleus of some galaxies are silent also. These facts may confirm the new picture since the "explosions" and other locally fast phenomena become slow ones in terms of $t$. The such "delayed explosions", continuing hours and days are really observed from time to time practically for all the such objects. Although the mechanisms of these phenomena may be complicated, but the dilated explosions also probably contribute to the total effect. Moreover, the such dilations may be used for the measuring the time dilation coefficient at the surface of the such compact objects.

9. New opportunities will be opened by GR for the theories of supernovae explosions and for understanding of the nucleosynthesis which may occurs not only at explosions of a supernovae, but also during longer time in the gravitationally-frozen superdense stars by escaping out at the explosions due to the phase transitions.

## 4 Conclusion

In GR the equations of motion and the identity $r(t) = r(\tau)$ along worldlines lead to the constraint for the proper times $\tau = \tau(t)$. As the result, at any $t < \infty$, one has $\tau(t) < \tau_g$, i.e. the constraint (7) restricts the proper time's moments by the world time moments on the globally defined hypersurfaces of simultaneity. This means that the proper times of falling particles become frozen in the contracting star so that at any finite coordinate time moment $t$ the proper time dilation is maximal at the centre and minimal on the surface of the star.

Thus, at any finite moment of the coordinate time GR predicts the gravitationally frozen superdense and supermassive stars with some internal structure where the space-time interval for matter is time-like (light-like for massless particles) and regular in entire space. The gravitationally-frozen matter is a new phase state of matter having new unusual and observable properties [3].

In contrast with the black holes as fully isolated objects, appearing according to GR only at $t \to \infty$, the predicting frozen objects are observable at any finite moment of the coorduinate time and may be sufficiently active. During the very slowered contraction the frozen star radiates (with high redshift), and at the internal phase transitions of matter with large energy liberation, may be observed the bursts and explosions.

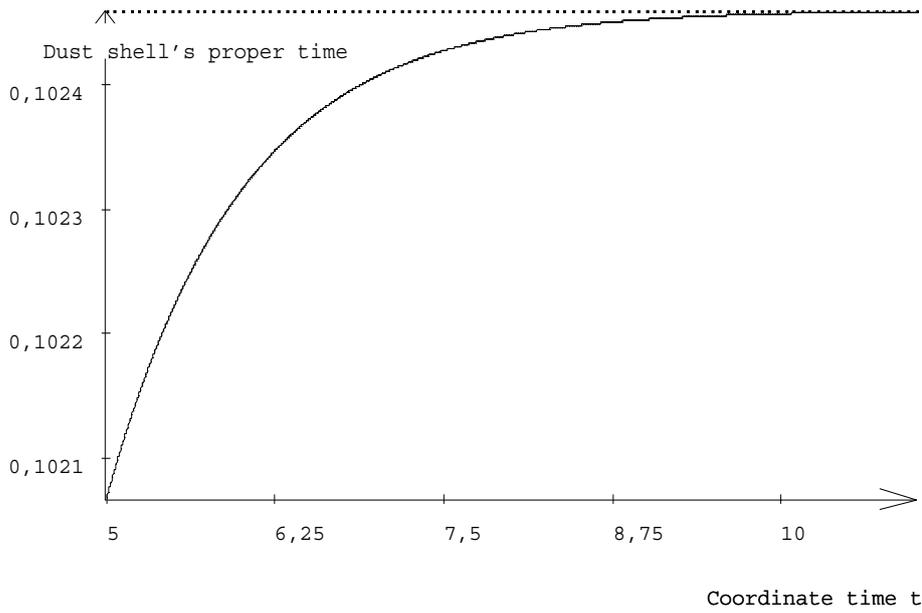

Fig.1. The constrained proper time $\tau(t)$ of a freely falling dust shell. The identity $r(t)=r(\tau)$ relates two kind of times and gives the invariant restriction $\tau[r(t)]<\tau_g$ for any finite $t$ ($\tau_g$ – dotted line). For the global set of synchronized coordinate clocks the intervals $\Delta t$ are equidistant, corresponding local proper times are *dilated* $\tau[r(t)]<t$. The initial values $t(r_1)=\tau(r_1)=0$ are taken at $r_1=1.1\ r_g$, times are in units $r_g/c$.